\documentclass[pdflatex,sn-nature]{sn-jnl}

\usepackage{graphicx}%
\usepackage{multirow}%
\usepackage{amsmath,amssymb,amsfonts}%
\usepackage{amsthm}%
\usepackage{mathrsfs}%
\usepackage[title]{appendix}%
\usepackage{xcolor}%
\usepackage{textcomp}%
\usepackage{manyfoot}%
\usepackage{booktabs}%
\usepackage{algorithm}%
\usepackage{algorithmicx}%
\usepackage{algpseudocode}%
\usepackage{listings}%

\usepackage{acronym}
\usepackage{siunitx}
\usepackage{lmodern}

\acrodef{EEG}{electroencephalography}
\acrodef{EMG}{electromyography}
\acrodef{MEG}{magnetoencephalography}
\acrodef{HMP}{hidden multivariate pattern}
\acrodef{PCA}{principal component analysis}
\acrodef{ICA}{individual component analysis}
\acrodef{S4}{structured state space sequence}
\acrodef{SSM}{state space model}
\acrodef{fMRI}{functional magnetic resonance imaging}
\acrodef{RT}{reaction time}
\acrodef{SVM}{support vector machine}
\acrodef{HMM}{hidden markov model}
\acrodef{RNN}{recurrent neural network}
\acrodef{BCI}{brain-computer interface}
\acrodef{SAT}{speed-accuracy trade-off}
\acrodef{ML}{machine learning}
\acrodef{AI}{artificial intelligence}
\acrodef{ACP}{average confirmation probability}
\acrodef{LSTM}{long short-term memory}

\graphicspath{/img}

\begin{document}

\title[Sequence models for by-trial decoding of cognitive strategies from neural data]{Sequence models for by-trial decoding of cognitive strategies from neural data}

\author*[1]{\fnm{Rick} \spfx{den} \sur{Otter}}\email{r.denotter@uu.nl, rickdotyt@gmail.com}

\author[1,2]{\fnm{Gabriel} \sur{Weindel}}

\author[1]{\fnm{Sjoerd} \sur{Stuit}}

\author[1]{\fnm{Leendert} \spfx{van} \sur{Maanen}}

\affil[1]{\orgdiv{Experimental Psychology, Helmholtz Institute}, \orgname{Utrecht University}}
\affil[2]{\orgdiv{Bernoulli Institute for Mathematics, Computer Science and Artificial Intelligence}, \orgname{University of Groningen}}

\abstract{Understanding the sequence of cognitive operations that underlie decision-making is a fundamental challenge in cognitive neuroscience. Traditional approaches often rely on group-level statistics, which obscure trial-by-trial variations in cognitive strategies. In this study, we introduce a novel machine learning method that combines \ac*{HMP} analysis with a \ac*{S4} model to decode cognitive strategies from \ac*{EEG} data at the trial level. We apply this method to a decision-making task, where participants were instructed to prioritize either speed or accuracy in their responses. Our results reveal an additional cognitive operation, labeled \textit{Confirmation}, which seems to occur predominantly in the accuracy condition but also frequently in the speed condition. The modeled probability that this operation occurs is associated with higher probability of responding correctly as well as changes of mind, as indexed by \ac*{EMG} data. By successfully modeling cognitive operations at the trial level, we provide empirical evidence for dynamic variability in decision strategies, challenging the assumption of homogeneous cognitive processes within experimental conditions. Our approach shows the potential of sequence modeling in cognitive neuroscience to capture trial-level variability that is obscured by aggregate analyses. The introduced method offers a new way to detect and understand cognitive strategies in a data-driven manner, with implications for both theoretical research and practical applications in many fields.}

\maketitle

\section{Introduction}
In order to perform an action in reaction to their physical environment, an individual first has to obtain perceptual input, which is then translated to an action output via one or more intermediate cognitive operations. That this procedure exists and consists -- at least partially -- of a sequence of cognitive operations is well-established \cite{dondersSpeedMentalProcesses1868, andersonHowCanHuman2007, zylberbergHumanTuringMachine2011}.

Traditionally, the question of identifying cognitive operations is approached through mathematically descriptive models of behavior, where behavior (often in the form of reaction times) is predicted through a principled set of modules representing cognitive operations \cite{newellUnifiedTheoriesCognition1994, andersonHowCanHuman2007, lairdSoarCognitiveArchitecture2019}. Such models have been instrumental in our understanding of cognition, but are mostly used to derive group-level statistics from individual trials, instead of determining single-trial-level cognitive operations. However, at the single-trial-level, variation in behavioral measures such as \ac{RT} is observed, and this has been attributed to variation in cognitive processes \cite{maanenNeuralCorrelatesTrialtoTrial2011, jonesUnfalsifiabilityMutualTranslatability2014, turnerInformingCognitiveAbstractions2015}.

Dominant theories on cognitive processing posit that variability in behavior is caused by a homogeneous but noisy process \cite{luceResponseTimesTheir1986, ratcliffComparisonSequentialSampling2004, brownSimplestCompleteModel2008}. However, another possibility is that this process is heterogeneous, where differences in reaction time are caused by variation of the associated cognitive processes \cite{evansBiasHumanReasoning1989, evansRapidRespondingIncreases2005, dutilhPhaseTransitionModel2011, maanenHowAssessExistence2014}.

One way to formally investigate whether behavior is caused by heterogeneous processes is through cognitive architectures \cite{newellUnifiedTheoriesCognition1994, andersonHowCanHuman2007, lairdSoarCognitiveArchitecture2019}. Cognitive architectures typically assume a strictly serial sequence of processing stages. In a theory-driven way, this assumption could be used to construct the sequences that constitute the separate cognitive processes and explain the heterogeneity in behavior.
Cognitive architectures have been linked to neural mechanisms using \ac{fMRI} \cite{borstUsingModelbasedFunctional2013, fechnerStrategiesMemorybasedDecision2016, yangRelianceEpisodicVs2024b} and \ac{EEG}/\ac{MEG} \cite{borstStagesProcessingAssociative2013, borstDiscoveryProcessingStages2015, andersonDiscoveryProcessingStages2016, portolesCharacterizingSynchronyPatterns2018, berberyanEEGbasedIdentificationEvidence2020, berberyanDiscoveringBrainStages2021, vanmaanenDiscoveryInterpretationEvidence2021}. However, all these methods make explicit assumptions concerning cognitive operations. Specifically, they either assume that cognitive processing resembles a cognitive model, or use experimental manipulations that are assumed to affect only a certain specific operation, similar to Sternberg's method of additive factors \cite{sternbergDiscoveringMentalProcessing1998}. None of these approaches allows to detect cognitive operations in a data-driven way, directly identifying them in neural signals. A novel machine learning method called \ac{HMP} \cite{weindelTrialbytrialDetectionCognitive2024} (a generalization of \cite{andersonDiscoveryProcessingStages2016}) achieves this in \ac{EEG} data by estimating, at the condition level, a number of cognitive operations. Fitting an \ac{HMP} model provides, per-trial, a probability distribution for the activation peak of each detected cognitive operation. This approach has been successfully applied in a number of previous studies \cite{andersonDiscoveryProcessingStages2016, berberyanDiscoveringBrainStages2021, berberyanEEGbasedIdentificationEvidence2020, vanmaanenDiscoveryInterpretationEvidence2021, zhangEffectsProbeSimilarity2017, portolesCharacterizingSynchronyPatterns2018, weindelTrialbytrialDetectionCognitive2024}. However, \ac{HMP} assumes that all trials in a condition share the same number of cognitive operations. Therefore, \ac{HMP} by itself is unable to account for different sequences of cognitive operations in a given experimental condition. In short, there is no method that uses neural signals to determine the presence of a cognitive operation at trial level.

A behavioral effect that is difficult to examine at the trial level using traditional approaches is the \ac{SAT} \cite{heitzSpeedaccuracyTradeoffHistory2014}, a well-known phenomenon of human decision-making. The \ac{SAT} reflects a decrease in task performance when participants respond faster, suggesting dynamic re-allocation of cognitive resources based on task demand. Experimental tasks where participants are motivated to focus on either speed or accuracy, combined with advances in neuroimaging, have gotten us closer to understanding how the brain makes decisions under time pressure \cite{ratcliffModelingResponseTimes1998, forstmannStriatumPreSMAFacilitate2008, forstmannCorticostriatalConnectionsPredict2010, bogaczNeuralBasisSpeedaccuracy2010, maanenNeuralCorrelatesTrialtoTrial2011, wenzlaffNeuralCharacterizationSpeed2011, boehmTrialtrialFluctuationsCNV2014, vanmaanenStriatalActivationReflects2016, vanmaanenDiscoveryInterpretationEvidence2021, weindelAssessingModelbasedInferences2021}. However, it remains a challenge to assess whether a participant indeed followed the task instruction to respond either fast or accurately on every single trial. Behavioral measures such as \ac{RT} and performance can be used at a condition level to show the effect of speed and accuracy strategies, but since there are only subtle differences in behavioral measures between strategies, it is difficult to identify at trial level which strategy was applied (but see \cite{maanenNeuralCorrelatesTrialtoTrial2011, hoOptimalitySensoryProcessing2012, boehmTrialtrialFluctuationsCNV2014, turnerInformingCognitiveAbstractions2015}). Methods using \acp{HMM} \cite{dutilhPhaseTransitionModel2011, kucharskyHiddenMarkovModels2021, kunkelHierarchicalHiddenMarkov2021} classify trials based on the temporal dynamics of \ac{RT} and performance over trials, but typically do not use any other trial-specific information. These methods do consider individual variation on the trial level, but they either only capture the part of the variation that is visible in \ac{RT} and performance, or do not provide any information on what the applied strategy is like at the level of the brain.

A method to accurately identify sequences of cognitive operations - thereby identifying strategies as a source of heterogeneity in behavior - should consider individual variation, while using information from a neural source that is closely related to cognitive operations. We propose the usage of \ac{ML}, for its ability to learn from multivariate data (\ac{EEG}) and sequences \cite{lawhernEEGNetCompactConvolutional2018, kuntzelmanDeepLearningBasedMultivariatePattern2021, guMambaLinearTimeSequence2024}. While current applications of \ac{ML} in analyzing \ac{EEG} data focus mostly on multi-second spans of \ac{EEG} data for classification of sleep stages \cite{eldeleAttentionBasedDeepLearning2021}, emotions \cite{songEEGEmotionRecognition2020a}, or motor imagery \cite{zhaoCTNetConvolutionalTransformer2024}, we propose to classify the predicted sequences of cognitive operations. Since cognitive operations occur at a time scale of 10-100 ms \cite{newellUnifiedTheoriesCognition1994}, it is important for any model to not only be able to recognize patterns at this time scale, but also consider the sequential nature of cognitive operations. Instead of classifying a longer sequence of \ac{EEG} data, we therefore need to classify each time step.

\begin{figure*}
    \includegraphics[width=\textwidth]{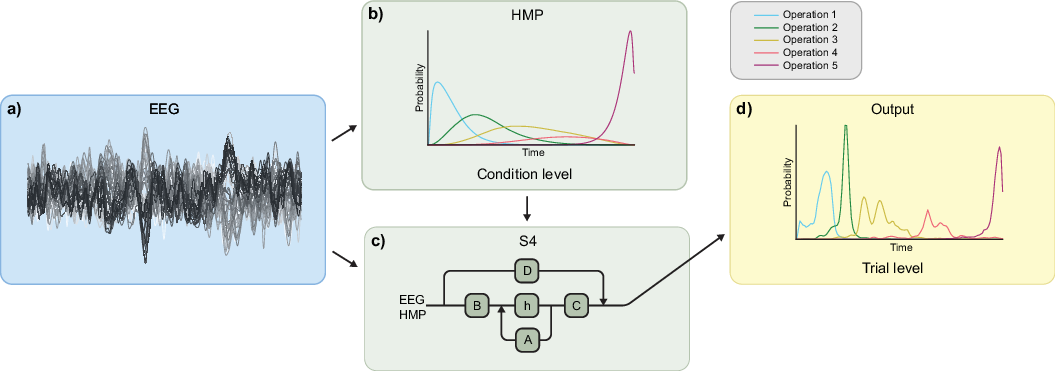}
    \caption{Conceptual representation of the introduced method. \textbf{a)}: \ac{EEG} data is fed into an \ac{HMP} model \textbf{b)}, which estimates cognitive operations per condition and outputs per-trial probabilities of the activation peak of each cognitive operation. \ac{HMP} probabilities \textbf{b)} and raw \ac{EEG} data \textbf{a)} are then processed using an \ac{S4} model \textbf{c)}, where $A$, $B$, $C$, and $D$ are parameter matrices with different effects on the current state $h$, based on their position in the model. \textbf{d)}: The \ac{S4} model outputs probabilities with loosened restrictions compared to \ac{HMP}, providing the probability of the onset of a cognitive operation at each time step. See text for details.}
    \label{fig:concept}
\end{figure*}

In the current work, we apply state-of-the-art machine learning methods to \ac{EEG} data (Figure \ref{fig:concept}). The goal is to detect patterns of neural activity associated with specific cognitive operations, which are expected to differ based on task demands, partly driven by the experimental manipulation. To achieve this, we apply an \ac{S4} model, chosen for its ability to model longer sequences and handle sequential dependencies \cite{guEfficientlyModelingLong2022, guMambaLinearTimeSequence2024}. We train the model to detect the onset of each cognitive operation using \ac{EEG} data and \ac{HMP} results as labels, which contain the per-trial probabilities of the activation peak of each cognitive operation. The trained model can then be used to investigate where and when differences in the neural signal occur, indicative of variation in cognitive operations. The results can be used to detect which cognitive operations were likely to have occurred at trial level, in turn showing the strategy that the participant likely followed. While previous work has studied how neural representations could be detected using machine learning \cite{lawhernEEGNetCompactConvolutional2018, guggenmosMultivariatePatternAnalysis2018, peelenTestingCognitiveTheories2023, jiangLargeBrainModel2024}, our work takes the sequential nature of cognition into account to understand, from perception to action, the end-to-end process of cognition on a by-trial basis.

Here, we applied this novel method to data from a \ac{SAT} task. In this task, participants were instructed to decide which of two sinusoidal gratings had the higher contrast. In a blocked design, participants were instructed to respond either accurately, or as fast as possible (see \cite{weindelMeasurementEstimationCognitive2021} for details). In addition to \ac{EEG} and behavioral data, \ac{EMG} data were also collected from muscles in the hands to index the intention to choose.

We were able to successfully recover the \ac{HMP}-estimated sequence of cognitive operations in held-out data. Moreover, we were able to isolate a cognitive operation that occurs \textit{only} on certain trials, but not others, and seems to be related to the task demand of being as accurate as possible. This event occurred more often under task instructions to be accurate than task instructions to be fast, and trials in which this event occurred were more likely to have been answered correctly by the participant. Finally, trials in which this event occurred were associated with changes of mind, as indexed in the \ac{EMG} information \cite{burleExecutiveControlSimon2002}. This is the first work that combined \ac{S4} and \ac{HMP} models to identify the sequence of cognitive operations on trial level, directly linking neural signals to cognitive strategies.

\section{Results}
\subsection{Determining the number of cognitive operations}
We fitted \ac{HMP} to our data to estimate the most likely position and number of events for both conditions. The \ac{HMP} fit resulted in 4 events in the accuracy condition, and 3 events in the speed condition. This suggests that an additional operation occurs in a significant portion of the trials in the accuracy condition, and that this additional operation does not occur in the majority of trials in the speed condition \cite{weindelTrialbytrialDetectionCognitive2024}. To make it easier to discuss these events, we label them according to their expected function, guided by timing in the trial and previous theory \cite{vanmaanenDiscoveryInterpretationEvidence2021, weindelTrialbytrialDetectionCognitive2024}. In order (for the speed condition), the events are labeled: \textit{Encoding}, \textit{Decision}, \textit{Response}. The additional operation in the accuracy condition is labeled \textit{Confirmation}, since we hypothesize that the operation functions as a confirmation of the outcome of the previous \textit{Decision} operation.
This result is consistent with previous \ac{HMP} decompositions of \ac{SAT} data \cite{vanmaanenDiscoveryInterpretationEvidence2021, weindelTrialbytrialDetectionCognitive2024}.
\begin{figure*}
    \includegraphics[width=\textwidth]{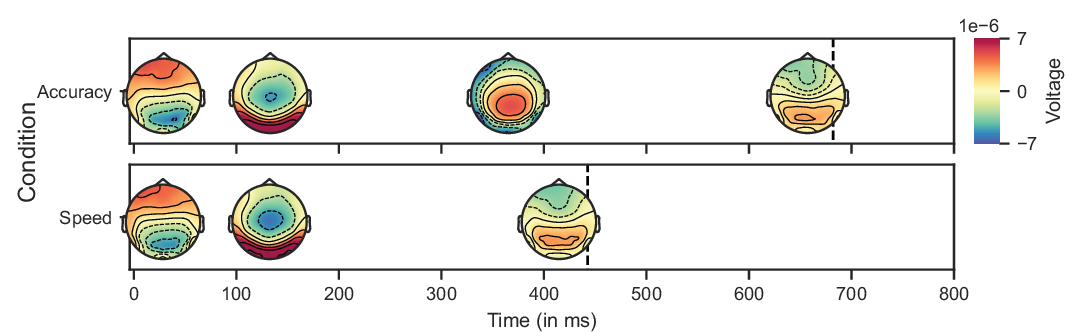}
    \caption{The result of the \ac{HMP} fit reveals an additional event in the accuracy condition that is not present in the speed condition. Operation topographies are aligned to the (on average) most likely point in time at which each activation peak occurs. The brain activity at this point is averaged over all trials and participants to create the content of the topographies.}
    \label{fig:fit}
\end{figure*}

\subsection{Decoding the sequence of cognitive operations}
\begin{figure*}[ht!]
    \includegraphics[width=\textwidth]{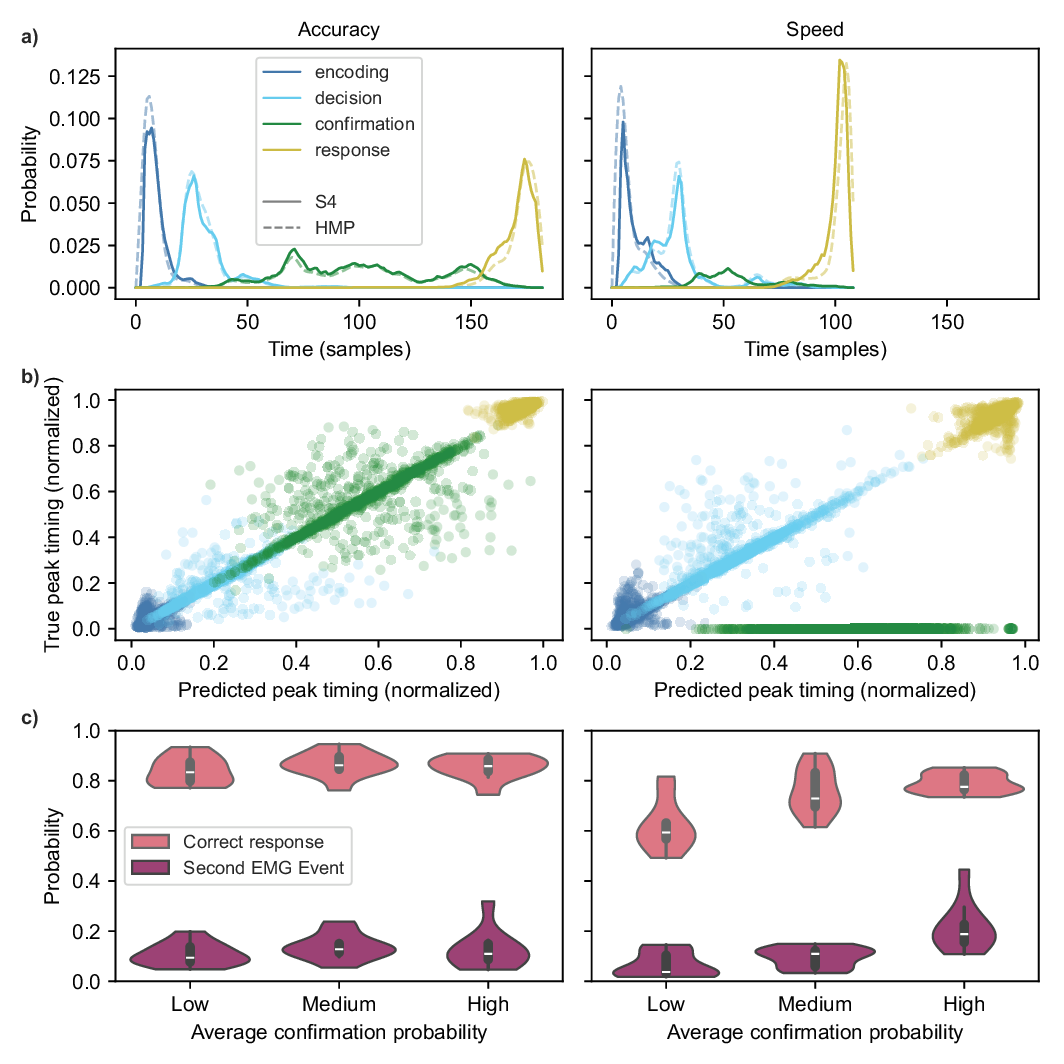}
    \caption{\textbf{a)} The model (solid lines) predicts \ac{HMP} probabilities (dashed line) well at single trial level. Shown here is one representative trial for each condition. \textbf{b)} An aggregate measure of true peak timing (Y-axis) and predicted peak timing (X-axis), values closer to the diagonal indicate a better prediction. \textbf{c)} The probability of correct response (Y-axis) increases with \ac{ACP} (X-axis). Probability of second \ac{EMG} event also increases with \ac{ACP}. Data split into tertiles based on \ac{ACP} for visualization purposes.}
    \label{fig:results}
\end{figure*}
We trained the \ac{S4} model on normalized \ac{EEG} data, with the \ac{HMP} probabilities as labels. The model minimizes the Kullback-Leibler divergence between predictions and true labels. The \ac{S4} model is able to learn the relationship between \ac{EEG} data and \ac{HMP} probabilities well. Panels a) and b) of Figure \ref{fig:results} show the performance of the model at a trial and aggregate level. These results indicate that the model's predictions align closely with \ac{HMP} labels, with generally only small deviations. Interestingly, the model does deviate from \ac{HMP} probabilities for the \textit{Confirmation} operation, which is predicted to be present in speed trials. This may indicate that participants do execute the \textit{Confirmation} operation on some trials, possibly indicating that their strategy differed from the experimentally intended strategy.

To understand the functional role of the \textit{Confirmation} operation, we compared \acf{ACP} to behavioral performance in the experiment. \ac{ACP} is the sum value of \textit{Confirmation} predictions, divided by the \ac{RT} and \textit{z} scored across all unseen data. We interpret this as the evidence that the \textit{Confirmation} operation occurs on a trial, relative to other trials. Using a generalized linear mixed model analysis, we found a significant main effect of time pressure condition on probability of correct response, indicating lower odds of correct response in the speed condition when compared to accuracy ($\beta = -0.56$, $SE = 0.04$, $z = -14.00$, $p < 0.001$, $OR = 0.57$, $95\%\,CI\,[0.53, 0.62]$, Figure \ref{fig:results}c). Importantly, a significant main effect of \ac{ACP} was found, with higher \ac{ACP} predicting increased probability of correct response ($\beta = 0.13$, $SE = 0.03$, $z = 3.76$, $p < 0.001$, $OR = 1.13$, $95\%\,CI\,[1.06, 1.21]$). Additionally, there was a significant interaction between time pressure condition and \ac{ACP} ($\beta = 0.24$, $SE = 0.04$, $z = 5.94$, $p < 0.001$, $OR = 1.27$, $95\%\,CI\,[1.17, 1.37]$), this interaction shows that the positive association between \ac{ACP} and probability of correct response was significantly stronger in the speed condition than in the accuracy condition. To summarize, \ac{ACP} was correlated to behavioral performance in both conditions, but more so when speeded responses were required. We applied this method to a second, smaller dataset on the \ac{SAT}, and found similar results (see Supplementary Section \ref{sec:boehm}).

We interpret significant bursts of \ac{EMG} activity as the intention to press a button \cite{burleExecutiveControlSimon2002}. We applied a generalized linear mixed model analysis to this information, with the outcome variable of observing either one or two \ac{EMG} events. We found a main effect of time pressure condition on \ac{EMG} activity, indicating that there were significantly lower odds to observe a second \ac{EMG} event in the speed condition ($\beta = 0.29$, $SE = 0.05$, $z = 6.00$, $p < 0.001$, $OR = 1.34$, $95\%\,CI\,[1.22, 1.47]$, Figure \ref{fig:results}c). Additionally, we observed a clear independent effect of \ac{ACP} on \ac{EMG} activity ($\beta = 0.28$, $SE = 0.04$, $z = 7.23$, $p < 0.001$, $OR = 1.32$, $95\%\,CI\,[1.23, 1.43]$). We also observe a significant interaction, where in the speed condition increases in \ac{ACP} increase the odds of a second \ac{EMG} event ($\beta = 0.30$, $SE = 0.05$, $z = 6.41$, $p < 0.001$, $OR = 1.36$, $95\%\,CI\,[1.24, 1.49]$). These findings indicate that the effect of \ac{ACP} on \ac{EMG} is consistent across conditions but was more strongly present in the speed condition.

\subsection{Further evidence for predicted confirmation}
\begin{figure*}[ht!]
    \includegraphics[width=\textwidth]{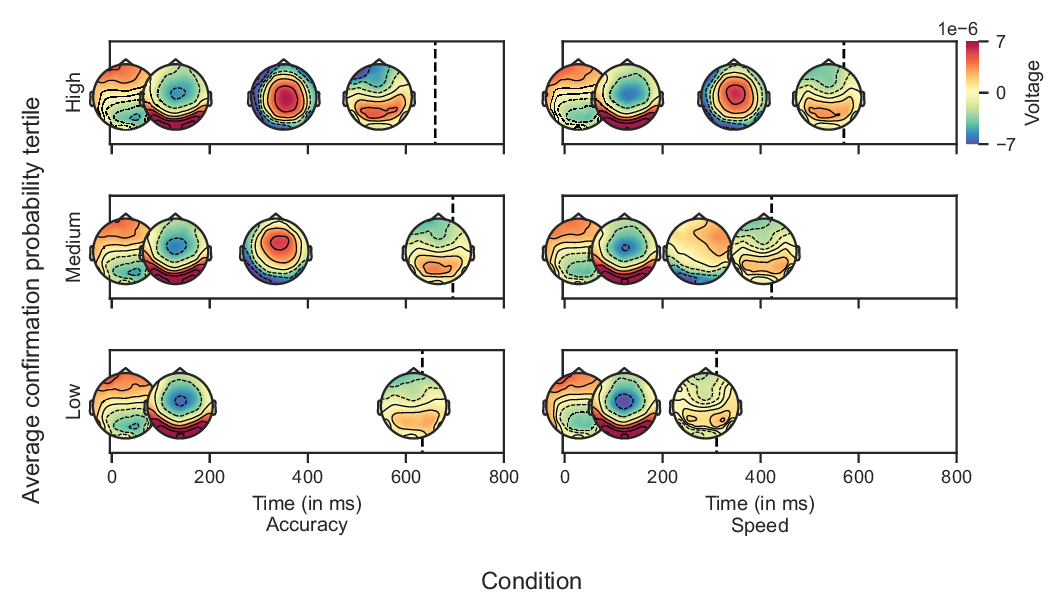}
    \caption{\ac{HMP} re-fit based on \ac{ACP} tertiles, in accuracy trials with low \ac{ACP}, the \textit{Confirmation} operation disappears, while in speed trials with high \ac{ACP}, the \textit{Confirmation} operation appears. These findings oppose the initial fit, indicating that trials in these subsets were more likely to have followed the opposite strategy.}
    \label{fig:hmp-refit}
\end{figure*}

Having determined the likelihood that an extra operation occurred at trial level, we can now use \ac{HMP} to validate these results. We calculated tertiles for \ac{ACP} values within condition and participant, and then split the data into 2x3 subsets on condition and \ac{ACP} tertiles. We then fitted \ac{HMP} to each subset, using the same fitting parameters and \ac{PCA} weights as the original fit, and fixing the parameters of the first event to those estimated during the initial fit since these are expected to be equal across conditions. As seen in Figure \ref{fig:hmp-refit} In the speed/high tertile, we find the additional operation that we previously only saw in the accuracy condition. Interestingly, we also lose the additional operation in the accuracy/low tertile. These results indicate that in the majority of the trials within these subsets, the participants were applying a strategy associated with the opposite task instruction.

\section{Discussion}
In this study we applied a combination of state of the art machine learning methods to decode cognitive strategies from neuroimaging data. While previous models were only able to classify strategies at condition level, or use simpler behavioral measures to classify strategy, our model was able to decode sequences of cognitive operations -- that we interpret as strategies -- from \ac{EEG} data at trial level, in held-out data. We applied the model to investigate the neural differences between speeded and accurate decision-making. By teaching an \ac{S4} model what patterns in \ac{EEG} data constitute which cognitive operations, we found that the additional \textit{Confirmation} operation in the accuracy condition is also predicted to occur in trials in the speed condition, with varying amounts of confidence. The model's predictions can be used to detect on which trials a participant was more likely to have followed a different strategy than experimentally intended. Specifically, we found that trials where the \textit{Confirmation} operation is predicted to occur, are more likely to have been answered correctly, especially in the speeded condition. Additionally, we found that trials where the \textit{Confirmation} operation is predicted to occur were more likely to contain an additional \ac{EMG} event, associated with changes of mind or confirmation of decision. The results were confirmed in a second \ac{SAT} dataset that differed with respect to the stimulus shown, using the same architecture, by re-training on the main dataset using the electrodes common between datasets (see Supplementary Section \ref{sec:boehm}).

Our findings have theoretical implications for understanding the selection of cognitive strategies in decision-making. By successfully modeling cognitive operations at trial level from \ac{EEG} data, we offer empirical support for dynamic variability in decision strategies, challenging the assumption that cognitive processes are homogeneous within conditions. Relating the model's predictions of the presence of an additional operation to behavioral outcomes (accuracy) and physiological markers of intent (\ac{EMG}) supports the hypothesis that individuals do not always perform as intended by the experimenter. This highlights the need for models of cognition to account for heterogeneity within conditions. Consequently, our approach demonstrates the potential of sequence modeling in cognitive neuroscience to capture variability that is often obscured by traditional analyses at aggregate level.
  
While these results provide useful insights into speeded decision-making, the interpretation of these findings requires the consideration of limitations inherent to the methodology and dataset. Training labels derived from \ac{HMP} estimations as probability distributions over time, do not represent definitive ground truth because \ac{HMP} probabilities are themselves estimates, not known labels. Moreover, \ac{HMP} probabilities sum to 1.0 over time, implicitly assuming all cognitive operations occur in each trial of a condition. We hypothesize that the \textit{Confirmation} operation occasionally occurs in speed trials despite labels assigning it zero probability, causing correct model predictions to appear as false positives under standard metrics. Similar challenges arise in incomplete annotations and positive-unlabeled learning \cite{bekkerLearningPositiveUnlabeled2020, zhaoPositiveUnlabeledLearningCell2021, kaliuzhnyiReducingEffectIncomplete2023}, where class absence is uncertain and addressed by adjusting loss functions. Our use of Kullback-Leibler divergence, an asymmetrical loss, allowed for the flexibility needed to predict incomplete annotations. The issue of false positive predictions complicates optimization and evaluation. Despite this issue, our conclusions remain valid since the \textit{Confirmation} operation predominantly occurs in accuracy trials and infrequently in speed trials, aligning with observed differences in \acf{ACP}. The model provides a lower bound of performance, meaning that labeling improvements could further enhance strategy detection.

Future work could explore contexts where \ac{HMP}-provided labels more closely approximate the ground truth, utilizing hyperparameter tuning and data transforms to enhance model performance. Many cognitive tasks currently lack trial-level clarity regarding the cognitive strategies used, creating opportunities for our model. For example, cognitive arithmetic involves distinct neural patterns for fact retrieval versus procedural strategies \cite{ashcraftCognitiveArithmeticReview1992, grabnerRetrieveCalculateLeft2009, groenewegHiddenSemiMarkovModel2021}, yet relies heavily on inaccurate self-reporting \cite{nisbettTellingMoreWe1977} for strategy identification. Our method could objectively classify trial-level strategies by modeling sequences of cognitive operations. This approach holds promise for numerous applications, from clinical diagnostics—such as identifying strategy abnormalities associated with psychiatric and neurological disorders \cite{maiaReinforcementLearningModels2011, huysComputationalPsychiatryBridge2016} like gambling disorder \cite{potenzaGamblingDisorder2019}—to validating therapeutic interventions. Additionally, our method could be beneficial in training complex skills in domains like aviation or professional sports, where recognizing deviations from an optimal sequence of cognitive operations could enhance instruction effectiveness.

\section{Conclusion}
In this paper, our goal was to introduce a novel method for the analysis of sequences of cognitive operations, or strategies. Furthermore, we applied the method to the problem of detecting strategies in a decision-making experiment. We found that an additional operation, estimated by \ac{HMP} to occur in the accuracy condition, but not in the speed condition, sometimes also occurs in the speed condition. These findings indicate that we can identify the executed strategy on a trial-by-trial basis from \ac{EEG} data. The success of this method implies possibilities in detecting differences in strategy at trial level in other areas of cognitive neuroscience as well, where this distinction is usually made based on experimental instruction, or self-reporting. Our method offers a way to detect strategies in a bottom-up manner. All code written for this project is available at \href{https://github.com/rickdott/SoCOM}{https://github.com/rickdott/SoCOM}.

\section{Methods}
\subsection{Experiment design}
We reanalyzed data from Weindel et al. \cite{weindelMeasurementEstimationCognitive2021}.
The data was collected from 20 participants (12 female, 8 male). Participants had to decide which of two concurrently presented sinusoidal gratings had higher contrast by pressing a button that triggered only after a configured force threshold was exceeded. 
The contrast was manipulated per-trial by selecting one of three contrast levels: 23, 51 and 93\%.
The difference was kept constant at 14\%. The speed-accuracy manipulation consisted of per-block instructions to focus on either Speed or Accuracy. The participants were told that speed required a mean reaction time near 400 ms and accuracy required a percentage of correct responses of 90\%, while maintaining reaction time below 800 ms. At the end of each block, participants received feedback about their mean reaction time and percentage of correct responses. Speed and accuracy instructions were swapped after every two 100-trial blocks. Finally, the force level was manipulated by defining high and low force values per participant representing respectively 2 and 20\% of their maximum voluntary force level. The force level was varied every three blocks. For more details, see \cite{weindelMeasurementEstimationCognitive2021}.

\subsection{Data collection and preprocessing}
The brain activity was recorded with a BioSemi Active II system (BioSemi Instrumentation, Amsterdam, the Netherland) using 64 \ac{EEG} electrodes mounted according to the 10-20 positioning system. Additionally, \ac{EMG} data was recorded from the flexor pollicis brevis muscle using a bi-polar montage of two electrodes on the thenar eminence of each hand. Both types of electrophysiological recordings were sampled at 1024 Hz.
\ac{EMG} information was preprocessed by defining different groups based on the number of peaks of muscle activity measured \citep[see][for the detection method]{weindelAssessingModelbasedInferences2021}. The response hand can be either Incorrect or Correct, depending on the expected response side. R indicates the response. Thus, IR/CR are situations where a single \ac{EMG} peak was recorded before the response, CCR/IIR is where two \ac{EMG} peaks in the same hand were recorded before the response, and ICR/CIR is when two \ac{EMG} peaks were detected, but the hand used switched between the two peaks.
The EEG data was preprocessed using the \textit{mne} Python module \cite{gramfortMEGEEGData2013}. First the data was re-referenced to the average of the electrodes. An \ac{ICA} was then applied to the data, band-pass filtered between 1 and 50 Hz for the \ac{ICA}, using the \textit{fastica} algorithm. Independent components were then compared with the vertical and horizontal electro-oculogram as given by the four electrodes set around the eyes (two at the outer canti, two below and above the left eye). Independent components with a \textit{z} score higher than 3.5 were marked for rejection. 
Each thereby identified artefactual component was removed from the unfiltered data (on average 1.25 components). Faulty electrodes (N = 1) were discarded during the average referencing and the \ac{ICA} computation and interpolated using spherical splines after \ac{ICA}. The data was then band-pass filtered between 0.01 and 50 Hz. Epochs up to 4 seconds after stimulus onset were then created. The epochs were baseline corrected using the 300 ms before stimulus onset and de-trended using the linear de-trending method implemented in \textit{mne}.

\subsection{Hidden multivariate pattern analysis}
To find the by-trial probability distributions of onsets of cognitive operations, we applied \ac{HMP} \cite{weindelTrialbytrialDetectionCognitive2024}. \ac{EEG} data were downsampled to 250 Hz and high-pass filtered at 1 Hz. First, we split the \ac{EEG} data into epochs, setting a lower limit for the reaction time to 200 ms, with no upper limit. We include 250 ms of additional samples before stimulus onset and after response for the temporal jitter transforms, but these samples are excluded in further \ac{HMP} analysis. We exclude trials where the force button malfunctioned by removing any trial where the applied force between stimulus and recorded \ac{RT} minus 200 ms exceeded a force threshold determined by subtracting one interquartile range from the median of the relevant force levels measured at all reaction times of the participant. On average, $42.05 (SD = 30.36)$ trials were removed per participant. Second, we perform \ac{PCA} on the variance-covariance matrix averaged over all the participants included in the training set, keeping 10 components, and we split the entire dataset into two subsets based on the speed/accuracy conditions. Third, we fit \ac{HMP} models on each subset. We use the cumulative fit method, which estimates parameters of each most likely event, given the previously estimated events. We determined the event width parameter of 45 ms by fitting \ac{HMP} models with an event width between 20 ms and 60 ms (steps of 5 ms) and chose the mean event width of the most stable \ac{HMP} solutions, determined by the number of estimated events. Finally, we label the resulting probability distributions based on theoretical assumptions about what occurs during each operation. The labels for the speed condition are: \textit{Encoding}, \textit{Decision}, \textit{Response}. For the accuracy condition, they are: \textit{Encoding}, \textit{Decision}, \textit{Confirmation}, \textit{Response}.

\subsection{S4 Model training}
We employed two different methods of splitting the data. For training the models that were used for behavioral comparisons, we applied an approximate participant-wise train/test/val split of 50/25/25\%, using a stratified sampling approach. We calculated per participant the difference between their average \ac{RT} and correct rate over speed/accuracy conditions. We then split the data so there was an even distribution of \ac{RT} and correct rate in each set. For the ablation study (see Supplementary Section \ref{sec:ablation}), the data was split using participant-wise k-fold cross validation (excluding participants from the validation and test sets). We apply median average deviation \textit{z} scoring to normalize all sets, using information from the training set only. Additionally, we add a \textit{Negative} class that is assigned a value which ensures that the class probabilities at each time step sum to $1.0$.

During training, start and end-jitter transforms are applied to each sample, selecting a new start point from the 250 ms of samples before stimulus onset, and a new end point from the 250 ms of samples after reaction time. This increases data diversity and decreases the predictive power of absolute time as a feature. These transforms were used to ensure that the model does not learn that a specific cognitive operation always occurs at a certain absolute time point, but rather that it occurs at a certain relative position within the trial. Without this addition, cognitive operations that are spatially similar and temporally adjacent (like our \textit{Confirmation} and \textit{Response} operations) are difficult for the model to discern. Additionally, including additional samples without any probability assigned to them teaches the model more about what specifically does not constitute the onset of a cognitive operation.

The model was trained using Kullback-Leibler divergence loss between the predicted probabilities and the \ac{HMP}-estimated probabilities. We used the NAdam \cite{dozatIncorporatingNesterovMomentum2016} optimizer with a learning rate set to $0.0001$. Training was stopped early if validation loss did not decrease for three subsequent epochs. The model weights for the epoch with the lowest validation loss were saved. Training was done on an NVIDIA A40 GPU with 12 GB VRAM made available. All code written for this project is available at \href{https://github.com/rickdott/SoCOM}{https://github.com/rickdott/SoCOM}.

\subsection{Model architecture}
\begin{figure*}[h]
    \includegraphics[width=\textwidth]{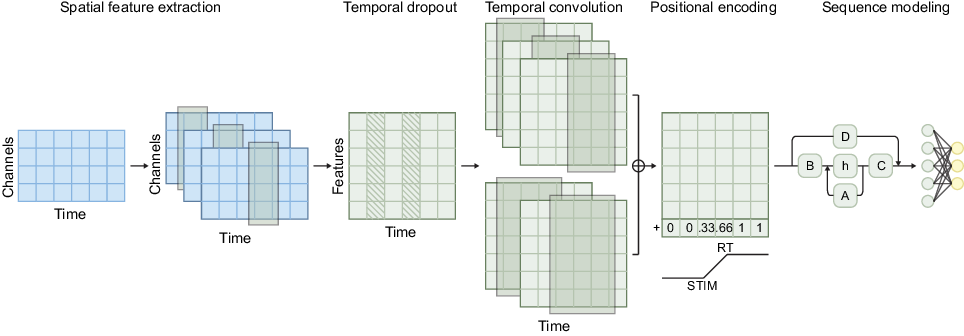}
    \caption{The model architecture used, blue indicates data, green indicates processing, yellow indicates output. First, spatial features are extracted from raw data, followed by temporal dropout. Then, temporal convolution is used to model temporal relationships at multiple time scales, after which positional encoding is added to the features. Finally, all features are fed into a Mamba sequence model and classifier.}
    \label{fig:architecture}
\end{figure*}

Our \ac{S4} modeling architecture uses the Mamba \cite{guMambaLinearTimeSequence2024} architecture. Since \ac{EEG} data is spatially informative, we first perform global spatial feature extraction using point-wise convolution. We iterate a $1 \times1 \times C$, kernel, where $C$ is the number of channels, over every time point (Figure \ref{fig:architecture}). By choosing a number of output channels greater than $C$, we ensure that the model can learn rich spatial relationships \cite{heDeepResidualLearning2015}. Temporal dropout is applied after spatial feature extraction to ensure that the model does not rely on specific time points. Next, all data is fed into the temporal feature extraction block, where a duo of convolutional layers is used to model temporal relationships at time scales of 3 and 9 samples, or 12 and 36 ms. The resulting features are concatenated. Since our trials vary in length, we add an additional feature containing a relative positional encoding vector, which contains zeroes before stimulus onset, an evenly spaced list of numbers in the range $[0, 1]$ starting at stimulus onset and ending at the reaction time, and ones after the reaction time. This helps the model know to which part of the trial each time step belongs. We do this to decrease the power of absolute time in the sequence as a feature, and increase the power of time relative to all other time points, which we expect to be more informative in this case. Then, we feed all features into 5 sequential Mamba \cite{guMambaLinearTimeSequence2024} layers, where temporal and spatial features are integrated across temporal contexts. Finally, a fully connected layer uses the resulting information to assign class probabilities to each time point. We validated the inclusion of parts of the architecture through an ablation study (see Supplementary Section \ref{sec:ablation}).

\newpage
\backmatter

\setcounter{section}{0}

\bmhead{Supplementary information}
\section{Applying method to another dataset} \label{sec:boehm}
\begin{figure*}[ht!]
    \includegraphics[width=\textwidth]{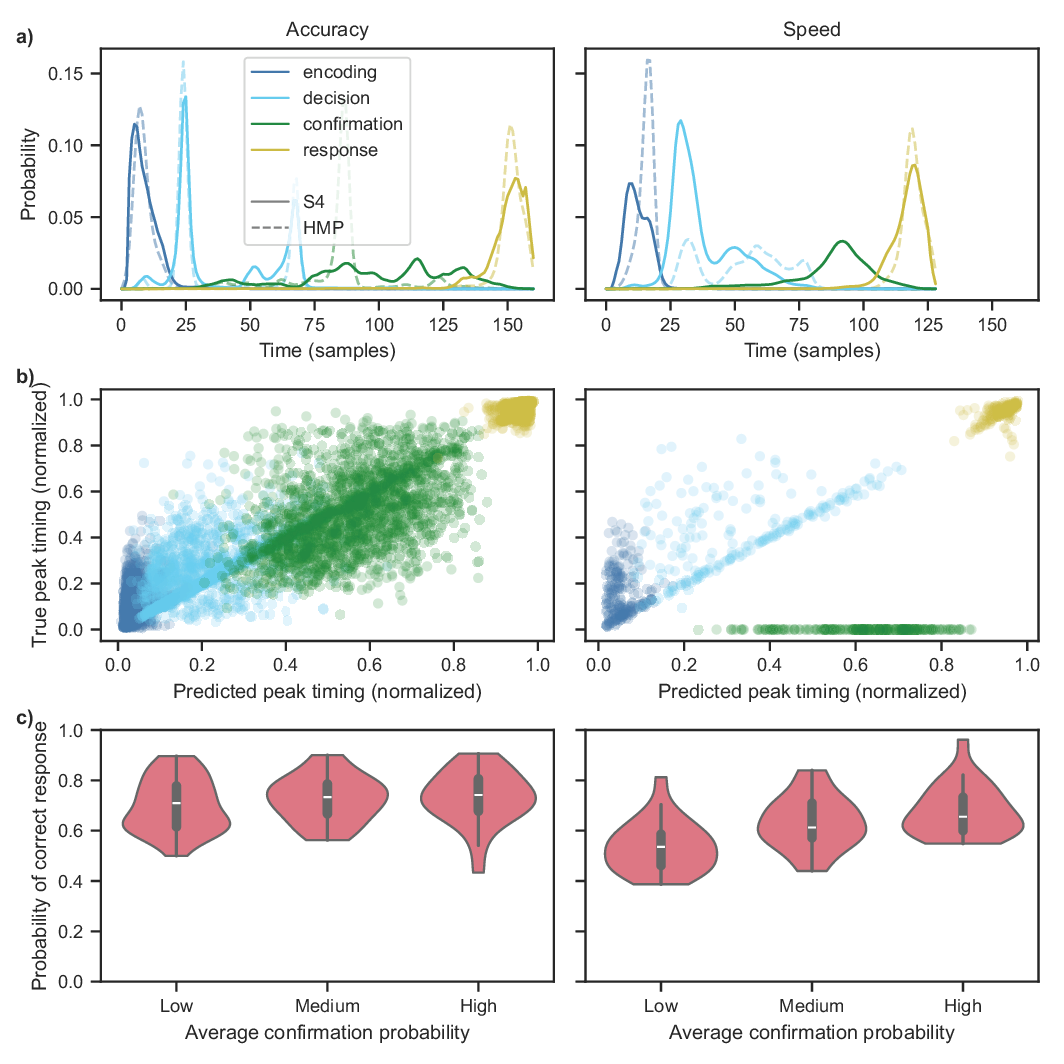}
    \caption{\textbf{a)} The model (solid lines) predicts \ac{HMP} probabilities (dashed line) well at single trial level. \textbf{b)} An aggregate measure of true peak timing (Y-axis) and predicted peak timing (X-axis), values closer to the diagonal indicate a better prediction. \textbf{c)} The probability of correct response (Y-axis) increases with \ac{ACP} (X-axis). Data split into tertiles based on \ac{ACP} for visualization purposes.}
    \label{fig:results_boehm}
\end{figure*}

When applied to another dataset on the \ac{SAT} \cite{boehmTrialtrialFluctuationsCNV2014}, we observe similar results (see Figure \ref{fig:results_boehm}). There were differences between the two datasets, namely that the second dataset had 25 participants (17 female), used 32 electrodes instead of 64, and did not record \ac{EMG} activity. The task was a random dot motion discrimination task, where participants had to decide whether a cloud of pseudo-randomly moving dots was moving to the left or to the right. Importantly, this study had 200 trials per participant, making it much smaller than the main dataset.

We fit \ac{HMP} to this dataset while fixing the number of expected events to 4 and 3 for the accuracy and speed conditions, respectively. We re-trained our original model on the dataset from the main experiment, using the common electrodes between datasets (all 32 from the second dataset). We applied the trained model to the entire second dataset and performed the same statistical analysis on \ac{ACP}, we found a significant main effect of time pressure condition on probability of correct response, indicating lower odds of correct response in the speed condition when compared to accuracy ($\beta = -0.35$, $SE = 0.07$, $z = -4.93$, $p < 0.001$, $OR = 0.71$, $95\%\,CI\,[0.61, 0.81]$, Figure \ref{fig:results_boehm}c). A significant main effect of \ac{ACP} was found, with higher \ac{ACP} predicting increased probability of correct response ($\beta = 0.14$, $SE = 0.06$, $z = 2.32$, $p < 0.05$, $OR = 1.16$, $95\%\,CI\,[1.02, 1.31]$). No interaction between time pressure condition and \ac{ACP} was found, indicating that the relationship between \ac{ACP} and probability of correct response was similar across conditions. We theorize that the difference in interaction strength could be caused by a ceiling effect in the accuracy condition in the main experiment.

\section{Model ablation study} \label{sec:ablation}
To validate the inclusion of each separate module of our model architecture, we performed an ablation study on the type and presence of spatial feature extraction, the number of convolutional layers, the presence of positional encoding, and the use of Mamba layers as opposed to a simpler \ac{LSTM} model. We used k-fold cross validation per participant and recorded the mean Kullback-Leibler divergence over each participant in the validation and test sets. As seen in Table \ref{tab:ablation}, a 1-D, or point-wise convolution outperforms not including spatial feature extraction and a linear layer. Increasing the number of convolutional layers (with 3, 9, and 27 kernel width respectively) past 2 convolutional layers did not improve performance. Removing the positional encoding vector decreased performance. Replacing the Mamba sequence model with an \ac{LSTM} worsened performance. Importantly, the variability in Kullback-Leibler divergence shows that we cannot conclusively say that for example a 1-D convolution layer outperforms a linear layer. Since optimizing the model architecture was not the goal of this paper, we simply went with the model that had the lowest average Kullback-Leibler divergence.

\begin{table}[ht!]
\centering
\begin{tabular}{@{}lllll@{}} \toprule
\textbf{Category} & \textbf{Model} & \textbf{KL divergence} & \textbf{Runtime (s)} & \textbf{Parameters} \\ \midrule
Spatial & No spatial & 1.13 (0.29) & 892.92 (262.48) & 2632777 \\
- & 1-D convolution & \textit{0.64} (0.13) & 833.22 (182.87) & 2690249 \\
- & Linear & 0.71 (0.31) & 815.95 (234.56) & 2690249 \\ \midrule
Temporal & 1 convolutional layer & 0.64 (0.13) & 833.22 (182.87) & 2690249 \\
- & 2 convolutional layers & \textit{0.61} (0.09) & 1760.86 (350.89) & 9662729 \\
- & 3 convolutional layers & 0.70 (0.15) & 2202.46 (408.67) & 21321033 \\ \midrule
Positional encoding & Present & \textit{0.61} (0.09) & 1760.86 (350.89) & 9662729 \\
- & Absent & 4.72 (0.40) & 1038.22 (358.80) & 21256581 \\ \midrule
Sequence model & Mamba & \textbf{\textit{0.61}} (0.09) & 1760.86 (350.89) & 9662729 \\
- & LSTM & 0.79 (0.58) & 2091.06 (788.84) & 24981473 \\ \bottomrule
\end{tabular}
\caption{Results of the model ablation study, cursive indicates best performance within category, bold indicates best performance overall, lower is better. Some models repeated for clarity. Numbers between parentheses are standard deviation. Runtime is averaged over all folds. KL = Kullback-Leibler.}
\label{tab:ablation}
\end{table}

\bmhead{Acknowledgments}
L.v.M. and G.W. have received funding from the European Union's Horizon 2020 research and innovation program under the Marie Skłodowska-Curie grant agreement No 101066503. L.v.M. is funded through the Air Force Research Laboratory grant No EOARD FA8655-22-1-7003.

\bmhead{Author contribution}
R.d.O., S.M.S., and L.v.M. initiated the project. R.d.O. implemented the method described, and wrote the first draft of the manuscript. G.W. provided assistance with the use of HMP, which he develops separately. S.M.S. and L.v.M. assisted with overarching guidance. G.W., S.M.S., and L.v.M. reviewed and edited the manuscript.

\end{document}